# Evaluating the Need and Effect of an Audience in a Virtual Reality Presentation Training Tool

Diego Monteiro [0000-0002-1570-3652], Hongji Li, Hai-Ning Liang [0000-0003-3600-8955], Yu Fu and Xian Wang

Xi'an Jiaotong-Liverpool University, Suzhou Jiangsu 15123, China
`haining.liang@xjtlu.edu.cn`

**Abstract.** Public speaking is an essential skill in everyone's professional or academic career. Nevertheless, honing this skill is often tricky because training in front of a mirror does not give feedback or inspire the same anxiety as presenting in front of an audience. Further, most people do not always have access to the place where the presentation will happen. In this research, we developed a Virtual Reality (VR) environment to assist in improving people's presentation skills. Our system uses 3D scanned people to create more realistic scenarios. We conducted a study with twelve participants who had no prior experience with VR. We validated our virtual environment by analyzing whether it was preferred to no VR system and accepted regardless of the existence of a virtual audience. Our results show that users overwhelmingly prefer to use the VR system as a tool to help them improve their public speaking skills than training in an empty environment. However, the preference for an audience is mixed.

## 1  Introduction

The fear of public speaking is the most mentioned in surveys on fears [1], and training for presentations can be challenging. However, in Virtual Reality (VR) environments, virtual characters can be to train presenter. Still, studies on the usefulness of a Virtual Audience (VA) for training people to make presentations are limited, focusing mostly on phobic people [2, 3]. However, in many situations, other people would like to practice interactive skills but lack the proper environment, as is the case in language learning [4, 5].

It has been shown in previous studies that people, regardless of their fear of public speaking, present different heart rates when exposed to VA [6]. However, this study was done with older technology, it is unclear if all people will be affected by a VA [7], because even simple changes as the viewing perspective can affect VR interaction [8, 9].

Further, even though much research has been done to demonstrate that a VA can yield positive results in the treatment of phobic people [10–13], not as much has been done to see the usefulness of a VA for the general population. To this end, in this research, we developed a VR environment to enable users to make virtual public speaking activities or presentations. To evaluate the acceptance of our system as a training tool, we gathered presenter's feedback.

## 2      Related Work

Virtual Reality Exposure Therapies (VRET) are therapies that use VR to expose patients to their fears [14]. Even before modern VR head-mounted displays (HMDs), VRET has shown potential [2, 11, 15, 16]. VRET, for instance, has been shown to work well for different kinds of anxieties [3, 17], such public speaking [18]. Further, one of the most prominent uses of VR is for training [19]. Several studies have shown that VR is a disruptive and useful tool for the training of various skills [16, 20, 21]; for technical [19], physical [22], and sociological training [23]. Even low-end VR can yield positive results [24]. However, more realism does not always present the best results [25].

### 2.1     VR as a Presentation Environment

If conditions are adequate, VR can elicit similar presence levels compared to the real world in an interview setting [26] it can even provoke fear [6]. Even though, these earlier models were not as realistic as what we can see in today's VR, people with social anxiety presented clear physical responses to them. However, these responses were not as prevalent in confident people. The reason for these responses may be because these participants thought the audience was not as realistic [6], or they ignored them like participants from a similar study [27]. Recent improvements in the realism level might have changed this paradigm or caused social anxiety people to get so used to the technology that they do not get triggered anymore. Hence, this raises a question (RQ1) *what are people's opinions about VA made with the current technology*?

A newer study analyzed if a realistic environment could increase the presence in a virtual presentation environment [27]. The researchers from this study created a classroom identical to the one available in their environment. They then asked participants to present to a VA and asked a semi-structured interview. They observed that participants overall appreciated the experience, and some felt it becoming more real as they got accustomed to it. However, some stated not perceiving the audience even though they appreciated the experience. These responses raise our main research question (RQ2): *is simply having a virtual classroom enough for some people?* To find the answer, we created a presentation environment in which participants could present either to a realistic empty room or to a partially 3D scanned VA.

## 3      The VR Presentation Environment

We developed a VR presentation environment to allow users to practice their presentations. We utilized pre-made models for the VA's body and movements. However, their heads were customizable 3D models of actual individuals; in our case the researchers (see Fig. 1). Based on the literature, we theorized that this would help the users feel more at ease before a real-life presentations [6, 27].

The VR environment contained a slide presentation on the back and footnotes in the front, which allowed users to present looking at the audience. The footnotes were akin to the presentation component in Microsoft PowerPoint; it allowed participants to have comments or texts in front of them (just like a prompter). The audience always gazed at the presenter.

## 4 Experiment Design

To evaluate whether the VR audience was indeed useful and to validate our VR presentation environment as a tool to practice presentations, the experiment had three conditions. In one condition, the participants practiced in front of a virtual audience (VRP). In another condition, the practice happened in an empty virtual room (VRE). As the control participants also practiced in the room alone without VR (RLE).

We made a series of presentations in English with topics that the volunteers were not familiar with according to their backgrounds. The participants prepared within a specific time to become familiar with the topic, and then presented the slides to a real-life audience. For all these later presentations, there were always the same audience members. Since the participants could not see their virtual face, the face model was not altered (see Fig. 1), and the participants' virtual hands matched their skin-tone.

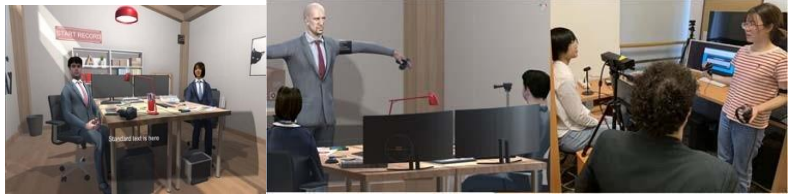

**Fig. 1.** The faces of the audience and the prompter (left). The virtual reality presenter and environment (middle). One presentation in real-life (right).

### 4.1 Environment Metrics

Each participant was asked to complete a simple questionnaire to collect demographic data. All participants were asked to self-assess their English level among the 6 levels of the CEFR [26]. After, they completed the Personal Report of Confidence as a Speaker (PRCS) [28], which is a true or false questionnaire with totaling 12 points. The more the volunteer scored, the more anxious the volunteer was supposed to be.

We used the 5-Point Likert Self-Statements During Public Speaking (SSPS) [29], which is referred to as a marker for short term treatments. The higher the score, the more confident the presenter. Further, we asked the participants to rate from 0 (worst) to 5 (best) how good they felt the conditions were, and their familiarity with the topic.

After going through the three conditions, we asked participants to rank each version according to a list of criteria and to answer a few open-ended questions (see Table 1).

**Table 1.** Open questions presented at the end of the experiment.

| ID | Question | Type |
|---|---|---|
| R1 | Rank your preparation after each version | Rank |
| R2 | Rank your nervousness after each version | Rank |
| R3 | Rank your version preference | Rank |
| Q1 | How do you feel about the virtual environment, not including the audience? | Open |
| Q2 | How do you feel regarding the virtual audience? | Open |
| Q3 | Is there anything you would change in the Virtual Environment? | Open |

### 4.2 Apparatus

We used an Oculus Rift CV1 as our HMD, and two standard 27" 4K monitor. We used the Oculus Touch to control the presentation even (the real-life presentation). Participants did not need to learn different controls for each version. We used monitors in real-life to simulate both the screen on the back and the text area facing the volunteer. All text size was calculated to be the same in VR and on the monitor.

### 4.3 Participants

We recruited a total of 12 participants (7 females; 4 males; 1 non-binary) from a local university. They had an average age of 19.88 (s.d. = 1.32), ranging between 19 and 23. No volunteer mentioned health issues, physical or otherwise. No participant was a native English speaker. Half had experience with VR systems before the experiment. The participants were not offered any reward to participate in the experiment.

### 4.4 Procedures

Each participant was assigned a specific order of the three conditions in which to practice the presentation. The order was Latin Square counterbalanced to mitigate carry-over effects. Participants were debriefed about the experiment.

Next, all participants were presented with a demonstration introduction of the Oculus Rift and a static virtual environment to get them acquainted with the VR HMD and set up—they were presented to the virtual room and a mock slide presentation.

Then, each participant was left alone to complete the training without being observed. The participants had 15 minutes to train for the presentation on the current condition. After this time, the participants were requested to remove the HMD and answer the questions. They then presented present the topic in real-life.

The participants were given 30 minutes to rest before starting the next condition. After all, conditions were tested; the last questionnaire was given to the participants. The participants were told that the open-ended questions could be answered in their native language if they felt more comfortable in doing so. In the end, the participants were thanked for their time, offered time to rest, and some refreshments. On average, each volunteer took one and a half hours to finish the whole data collection process.

## 5   RESULTS

The data were analyzed using both statistical inference methods and data visualizations. We conducted Mauchly's Test of Sphericity. We also employed Repeated Measures ANOVA (RM-ANOVA) using Bonferroni correction to detect significant main effects. If the assumption of sphericity was violated, we used the Greenhouse-Geisser correction to adjust the degrees of freedom. To control for both Type I and Type II errors we chose $p < 0.10$ for our analyses [30]. For the open-ended questions, we analyzed the sentences individually.

### 5.1   Pre-Questionnaire

We defined the bins for PRCS, normally distributed, as between 0 and 4 as not so nervous, until 8 as regular nervousness. Anything above was considered as extremely nervous. We dubbed these bins as Group Not Anxious (NA), Standard Anxious (SA), and Extremely Anxious (EA), respectively. 50% of the EA participants declared their level as B2. The other volunteers were mainly self-declared as B1.

### 5.2   Questionnaires of Nervousness and Preference Order

The SSPS questionnaire revealed significant differences after having practiced in the virtual environment the participants overall felt more positively to present to a real audience ($F_{2, 22} = 3.051$, $p = .068$). The positive effect was more present on the NA participants ($F_{2, 6} = 4.043$, $p = .077$). The post-hoc analysis revealed that VRP was especially more successful (see Fig 4) in making the participants feel less anxious ($p = .085$).

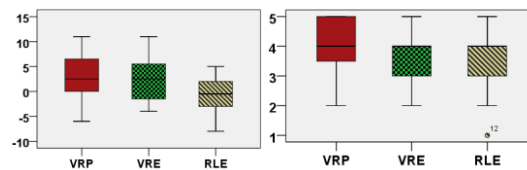

**Fig. 2.** SSPS answers after training in each version, VR versions score slightly better. Participants' ratings for each condition as a training tool.

The participants rated the VRP version the best as a training tool (see Fig 4). However, EA participants did not rate it more positively as a training tool than the other versions. The main difference in ratings came from NA participants.

The overall ranking disregarding PRCS can be seen on Fig 5. 75% of the EA participants chose one of the empty rooms as the version they felt most prepared on (R1).

However, they all selected the VRP as their second option. All EA participants selected the VRP as the version in which they felt the most nervous (R2). Half chose as their first choice, and 50% preferred it as their second version (R3); only one EA participant chose VRE as their first choice. Overall, VRP was the condition that received the most positive overall feedback. Only one NA participant chose VRP as their first choice. The SA participants consistently chose it as their first choice.

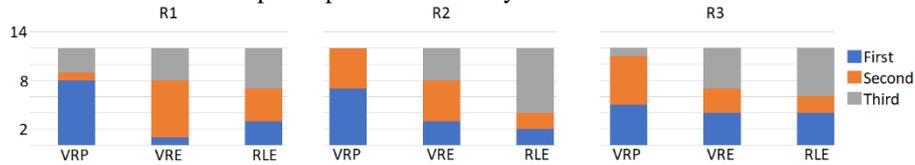

**Fig. 3.** Ranking to questions from R1 to R3, in their respective order from top to bottom.

### 5.3 Open-Ended Questions

The analysis of the interviews showed positive results and feedback. For Q1, one of the most common words reported by the volunteers is that the environment made them feel "good," and it was or felt "real".

Q2 and Q3 showed that participants believe that a virtual audience made them nervous, but they thought it was good. For example, P3 said "[I feel] a little bit nervous, especially when the audience's action changes." Some participants mentioned this as a positive point for making them better prepared to speak to the real audience later. P11 said that "There will be some tension, but it will then be easier in reality." P4 mentioned that the audience was good even though they made him feel "serious".

The EA participants were the ones who most used words such as nervous and scared. However, those were also the ones who commented most positively about the use of the application. As P3, who scored the maximum in the PRCS, puts it, "[It] made me not so relaxed but really help me improve my presentation." In general, participants said that they did not want to change the audience.

### 5.4 Discussion

The answers to the open questions and the ranking suggest a preference to VRP, which also indicate that the use of this VR tool for training has the potential to be well-accepted, and is aligned with previous research [6, 27]. Further the answers also show a positive aspect of having the faces of the real audience in the tool, which was described to give a "good nervousness". Hence, the participants opinion about a VA with the current technology (RQ1) is positive.

The rankings and the open-ended questions led us to believe that the fear factor generated by the virtual audience is favorable to the participants who, in return, train harder, or started feeling more secure with themselves. This is expected based on the exposure therapy treatments found in the literature [6]. This might indicate that an even greater audience might bring even better results.

Overall, VRP was verbally declared the most popular tool for a training presentation environment. Thus, in most cases, it is valuable to have a virtual audience interacting with the presenter. However, some people are satisfied with training in an empty room, which answers positively to the RQ2.

Although most participants are content with the VRP version, we recommend adding a toggle button. This should satisfy most users who might appreciate the choice. This may be the case especially because it seems the requirements for people who are EA are different than that of those who just need to practice presentation.

## 6      Conclusion

In this paper we explored whether there is a need (and the effect) of a virtual audience in a presentation training scenario using Virtual Reality (VR). We did this analysis through a series of subjective metrics which indicate that even though most people do appreciate the virtual audience generated with the current technology, and it makes them somewhat positively nervous. For some people, the current virtual audience is not necessary, they are just as satisfied with the system without it.

Further, we observed that a simple audience that follows the users by looking at them is enough for creating a stimulating training environment for presentations in a VR environment. This is a system that aims to invoke feelings of anxiety and promote effective training. This VR environment was able to make users nervous and rated 4 out of 5 as an effective training tool. We observed that the development of a training tool for public speaking training is the most adequate when it can have the audience and users would preferred to have the option to it on and off, depending on their preferences. Overall, the tool we developed seems to be able to help students.

## Acknowledgments

We would like to thank the participants for their time. This research was partially funded by XJTLU Key Program Special Fund and XJTLU Research Development Fund.